\documentclass[a4paper,11pt]{article}

\usepackage{pos}
\usepackage{hyperref}
\usepackage{lipsum} 

\title{ Search for high-energy neutrino emission from hard X-ray AGN with IceCube}

\ShortTitle{Neutrinos from Hard X-ray AGN}

\author{The IceCube Collaboration \\{\normalsize \normalfont(a complete list of authors can be found at the end of the proceedings)}\\}

\emailAdd{sgoswami2@crimson.ua.edu}
\emailAdd{jmsantander@ua.edu}

\abstract{

Active Galactic Nuclei (AGN) are powerful astronomical objects with very high luminosities. Theoretical arguments suggest that these objects are capable of accelerating particles to energies of 10$^{20}$ eV.  In environments with matter or photon targets, cosmic-ray interactions transpire leading to the production of pionic gamma rays and neutrinos. Since the AGN environment is rich in gas, dust and photons, they are promising candidate sources of high-energy astrophysical neutrinos. While the neutrinos manage to escape, the gamma rays may further interact and cascade down to hard X-rays in environments with sufficiently large photon or gas targets. We have used 12 years of IceCube data to perform a stacked search and a point source search for high-energy neutrino emission from hard X-ray AGN sampled from \textit{Swift}-BAT Spectroscopic Survey (BASS) and present the results of these two analyses. 

\vspace{4mm}
{\bfseries Corresponding authors:}
Sreetama Goswami$^{1*}$, Marcos Santander$^{1}$, James DeLaunay$^{1}$, George C. Privon$^{2}$\\
{$^{1}$ \itshape The University of Alabama}\\
{$^{2}$ \itshape NRAO}\\[4mm]
$^*$ Presenter

\ConferenceLogo{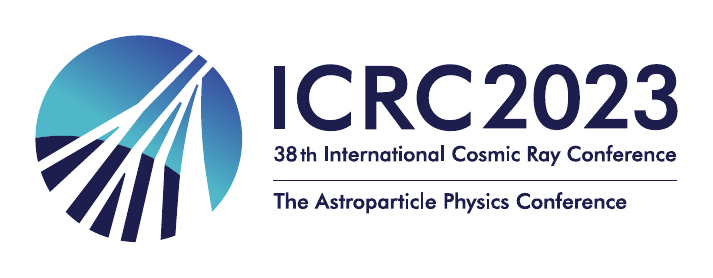}

\FullConference{The 38th International Cosmic Ray Conference (ICRC2023)\\ 26 July -- 3 August, 2023\\ Nagoya, Japan}
}

\begin{document}

\maketitle

\section{Introduction}\label{intro}

Active Galactic Nuclei (AGN) are one of the most luminous astronomical objects in the entire Universe emitting vast amounts of energy spread over a large range of frequencies \citep[see reviews for e.g.][]{doi:10.1146/annurev-astro-082214-122302,Padovani:2017zpf}. As the name suggests, these are the nuclei of active galaxies that are observed with extreme levels of radiation from the core region as compared to other regular galaxies. The basic structure of AGN consists of a supermassive black hole (SMBH) surrounded by an accretion disk. This central region is enveloped by a roughly torus-shaped structure rich in dust and gas known as the dusty torus. AGN may also be accompanied by a relativistic and highly collimated jet. The class of AGN with jets directed towards the Earth having an angle of less than $\sim 15^{\circ}$ with respect to the line of sight are blazars. The blazars are observed with a high luminosity that is believed to originate from relativistic beaming effects as the ionized plasma in the jets of blazars point towards the observer \cite{Urry_1995}.

AGN have the potential ability to accelerate charged particles to ultra-high  energies of about 10$^{20}$ eV \cite{Mbarek_2019}. There exist suggestions in the literature of AGN being promising sources of neutrinos \cite{PhysRevLett.66.2697,1997ApJ...488..669H,Murase:2022feu}. Within the AGN environment, the accelerated particles may undergo interactions and lead to a production of neutral and charged pions that upon decay produce neutrinos and gamma rays. There exists evidence of possible neutrino emission from individual AGN, i.e., TXS 0506+056 \cite{doi:10.1126/science.aat2890} and NGC 1068 \cite{doi:10.1126/science.abg3395}. However, searches aimed at finding neutrino emission from a class of sources bright in gamma rays have found no significant correlation \citep[see for e.g.,][]{2017ApJ...835...45A,Huber:2019lrm}. Theoretical models have thus suggested that the sources of high-energy neutrinos are possibly hidden and opaque to gamma rays \citep{Murase:2022feu}. The models propose that the pionic gamma rays produced in the AGN environments may further interact with ambient photons and cascade down to a flux of MeV gamma rays or hard X-rays \cite{Gao:2018mnu,Petropoulou_2020}. We perform two analyses to search for neutrino emissions from AGN that are observed in the hard X-ray regime of the 14 - 195 keV energy band.

\section{Neutrino data and the catalog}

The IceCube Neutrino Observatory is a neutrino telescope at the South Pole that uses a cubic kilometer of instrumented ice to detect and characterize astrophysical neutrinos \cite{Aartsen:2016nxy}. IceCube has reported the observation of diffuse fluxes of neutrinos that can be approximated by a single power-law function \cite{Aartsen_2019,Abbasi_2022}. We have chosen a neutrino dataset consisting of 12 years of muon neutrino tracks from 2008 to 2020 that uses the same selection criteria used for the dataset with  10 years of data as described in \citealp{PhysRevLett.124.051103}. These muon tracks constitute an all-sky neutrino dataset with an angular resolution of $< 1^{\circ}$ at  energies above 1 TeV \cite{Abbasi_2021}.

\begin{figure}
    \centering
    \includegraphics[width=\textwidth]{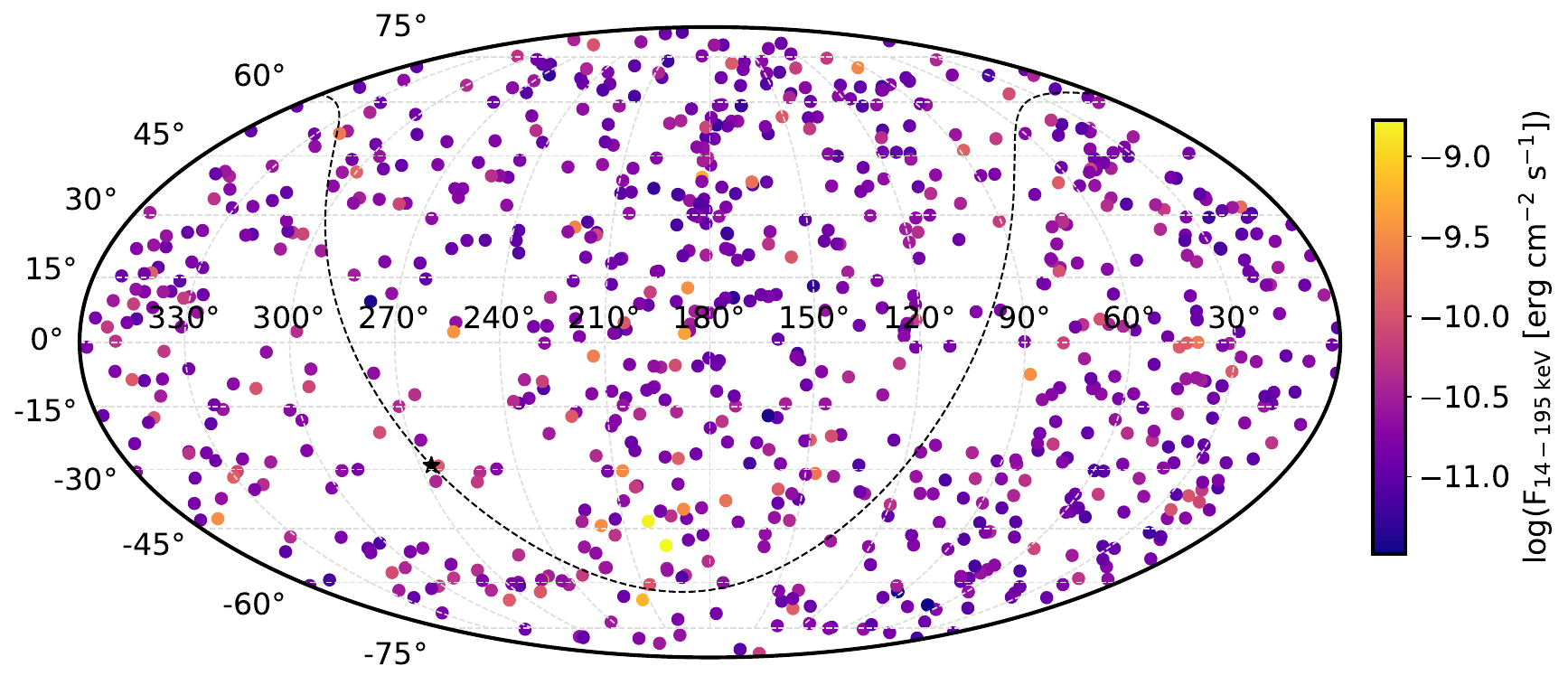}
    \caption{The  skymap in equatorial coordinates showing the position of BASS sources and the color of each source represents the intrinsic hard X-ray flux (14 - 195 keV). The black dashed line shows the galactic plane and the black marker shown as a star represents the galactic center in the image.}
    \label{fig:skymap}
\end{figure}

\textit{Swift}-BAT AGN Spectroscopic Survey\footnote{\href{https://www.bass-survey.com/}{The BASS project}}, or the BASS catalog \cite{2017ApJ...850...74K,Ricci_2017}, is a comprehensive selection of AGN compiled from the sources reported in the 70-month \textit{Swift}-BAT all-sky hard X-ray survey \cite{Baumgartner_2013}. It is an all-sky and the most complete list of AGN detected in the energy regime of 14 - 195 keV (see Figure \ref{fig:skymap}). It contains estimates of physical properties such as the distance, the intrinsic hard X-ray flux and the column density along the line of sight ($N_{\text{H}}$) and reports the type of AGN.  We have selected 836 AGN from the BASS catalog DR-1 for this work. There are two additional sources in the BASS catalog, SWIFT J1119.5+5132 and SWIFT J1313.6+3650A. They were excluded since the two sources did not have an estimate for the column density which is relevant to this study. 

\section{The Analyses}\label{sec:analyses}
We have performed two analyses to search for evidence of neutrinos emitted by AGN detected in hard X-rays described as follows.  The analyses are based on the statistical method of a time-integrated unbinned-likelihood hypothesis test \cite{ACHTERBERG2006282, BRAUN2008299}.

\begin{itemize}
    \item \textit{Stacked search}: In this analysis, we search for a correlation between neutrinos detected by IceCube and different classes of hard X-ray AGN. The AGN in the catalog can be classified as follows:  first, according to the type of AGN into two classes: (i) non-blazar AGN (732) and (ii) blazars (104), and second, according to the estimated value of column density into three classes, (i) unobscured sources with $N_{\text{H}}$ < 10$^{22}$ cm$^{-2}$ (457), (ii) obscured sources with 10$^{22}$ cm$^{-2}$ < $N_{\text{H}}$ < 10$^{24}$ cm$^{-2}$ (323) and (iii) Compton-thick sources with $N_{\text{H}}$ > 10$^{24}$ cm$^{-2}$ (56). We search for cumulative neutrino emissions from all the 836 AGN in our catalog and from each of the five classes mentioned above by stacking the sources in each class \cite[see for e.g.][]{Aartsen_2015}. It is assumed that the neutrino flux is correlated with the intrinsic hard X-ray flux and therefore, each source carries a theoretical weight proportional to the intrinsic hard X-ray flux. We use the values of intrinsic X-ray fluxe for the sources as reported in the BASS catalog where the authors correct the observed fluxes for absorption to obtain intrinsic fluxes \citep{Ricci_2017, 2017ApJ...850...74K}.  An additional case is tested using equal weighting for all AGN in the catalog. This is for comparison purposes with the other six hypotheses tested.
    
    \item \textit{Point Source Search}: In this analysis, we draw a shorter list of 43 sources that are the most promising sources of neutrinos based on their higher hard X-ray flux and location in the sky that is favorable to detection. We have searched for neutrino emissions from these individual point-like sources.
\end{itemize}

\section{Results}\label{sec:results}
\begin{figure}
    \centering
    \includegraphics[width= 0.45\textwidth]{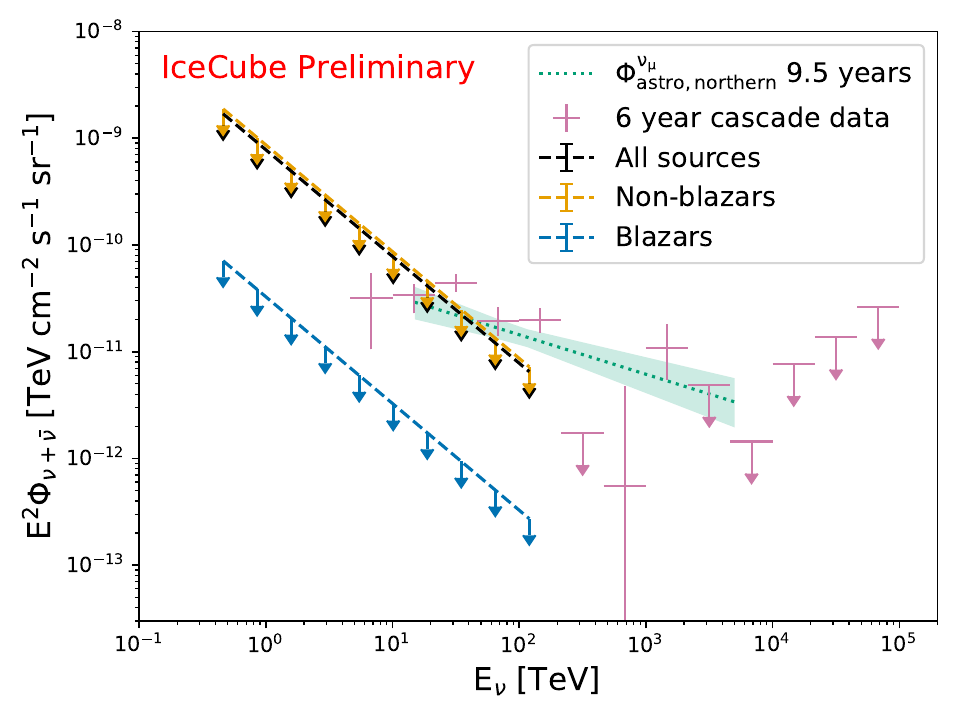}
    \includegraphics[width= 0.45\textwidth]{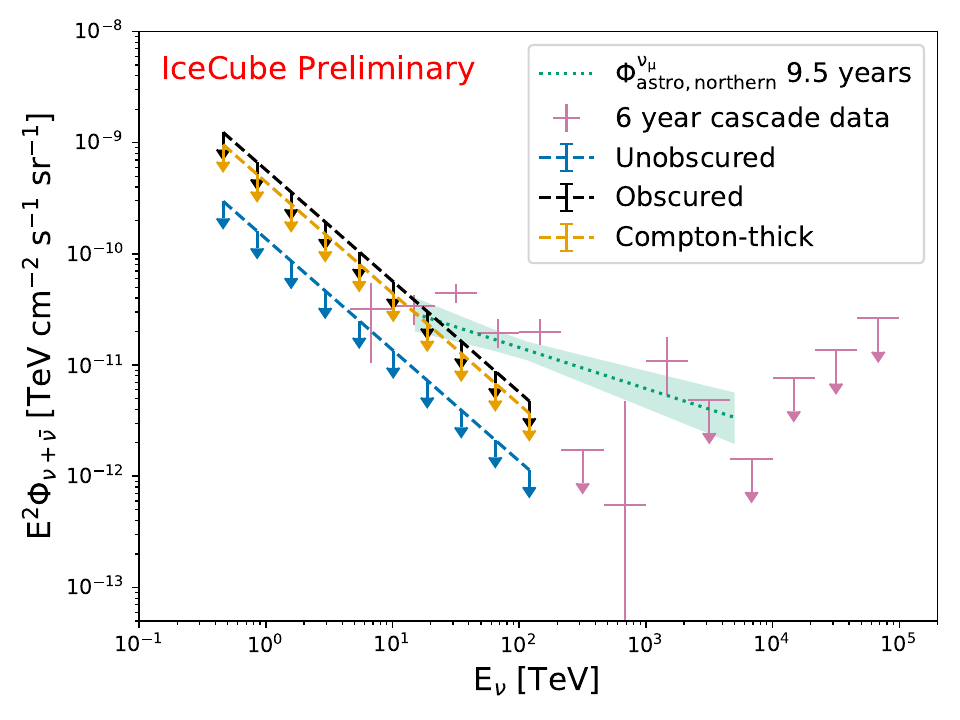}
    \caption{The figures show the 90\% CL flux upper limits obtained from the different AGN classes tested in the stacking analysis. Left: Right: The fluxes obtained from the analysis have been scaled to include the AGN that remain undetected and thus not in the current catalog. This scaling is done using catalog completeness factors computed for each class of AGN.  }
    \label{fig:uppperlimits}
\end{figure}

From the stacked search, we do not find evidence of significant emission from any of the seven hypotheses tested. Therefore, we have calculated the 90$\%$ confidence level (CL) flux upper limits for the source classes. In Figure \ref{fig:uppperlimits} (left), we show the flux upper limits obtained from all 836 AGN, the 732 non-blazar AGN and the 104 blazars and in Figure \ref{fig:uppperlimits} (right), we show the flux upper limits obtained from the 457 unobscured AGN, 323 obscured AGN and the 56 Compton-thick AGN. For comparison purposes, we have additionally shown the all-sky diffuse fluxes of neutrinos obtained by IceCube using the muon neutrino tracks in dotted light blue line \citep{Abbasi_2022} and using the cascades originating mostly from electron and tau neutrinos in pink \citep{Aartsen_2019}. The fluxes for each class of AGN in the figure are representative of all the AGN in the Universe belonging to that class after catalog completeness correction. The catalog completeness fraction for each AGN class is the ratio of neutrinos emitted by AGN in the catalog  and the total neutrino contribution from the entire source population. We have used luminosity functions from blazars  \cite{BlazarCatComp} and  non-blazar AGN  \cite{2014ApJ...786..104U}  and integrated them over all redshift to find the total neutrino flux.

From the point source search, the top two sources that result in the highest levels of significance are NGC 1068 and NGC 4151. A previous analysis reported the observation on NGC 1068~\cite{doi:10.1126/science.abg3395} and the results of this analysis are compatible with the evidence previously found.

\begin{figure}[h!]
    \centering
    \includegraphics[width= 0.45\textwidth]{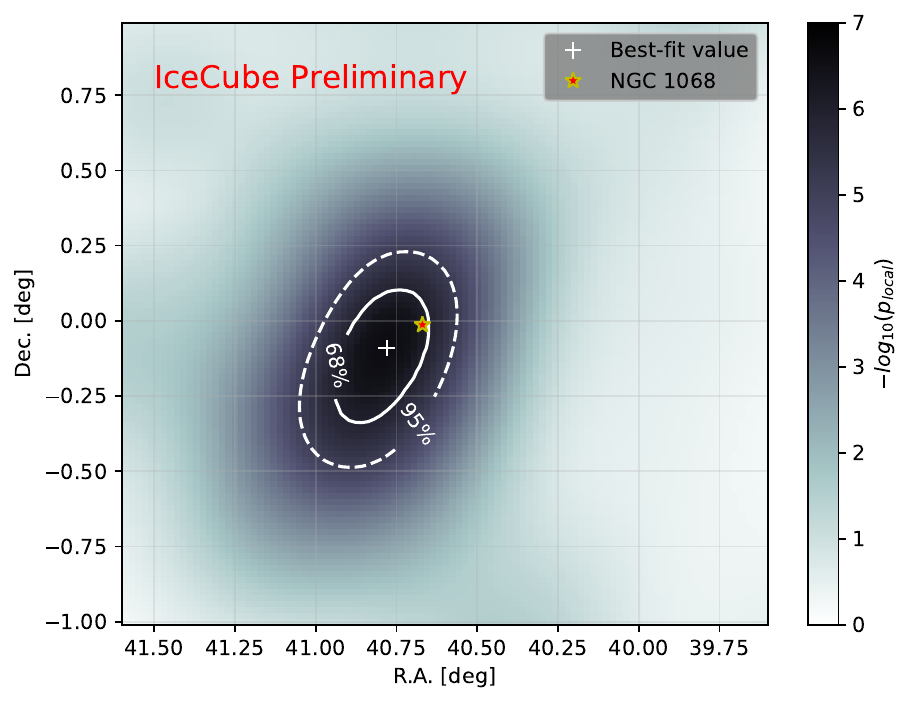}
    \includegraphics[width= 0.45\textwidth]{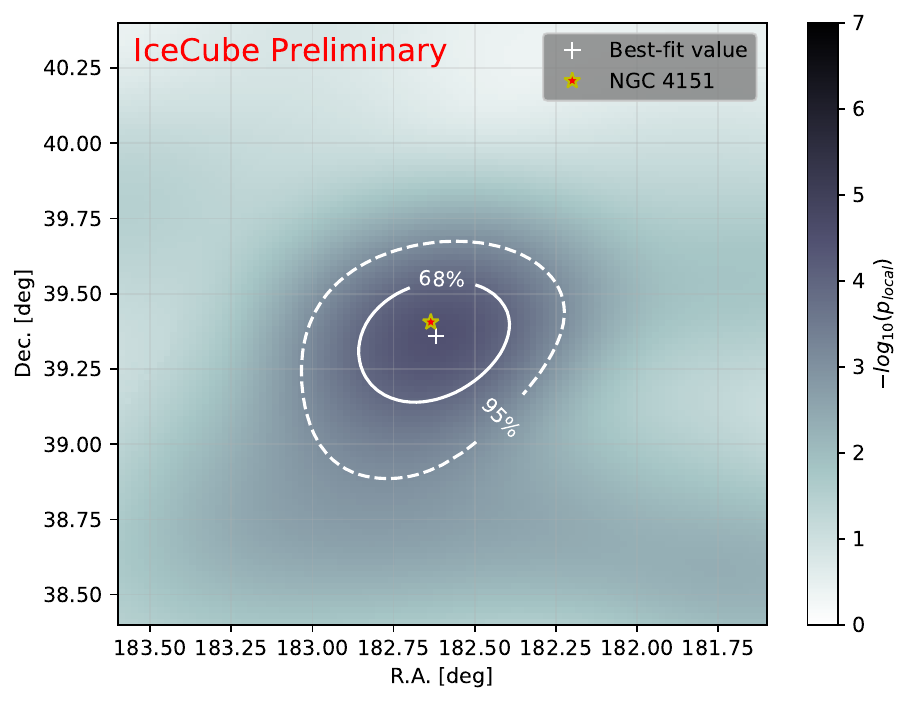}
    \caption{A scan of the region around the location of Right: NGC 4151 and Left: NGC 1068, showing the local $p$-values. The white cross shows the location of the best-fit position obtained from the point source search and the red star shows the location of the source as obtained from the catalog. The white unbroken and dashed lines show the 68\% and the 95\% contour lines, respectively, around the sources computed using Wilk's theorem \cite{10.1214/aoms/1177732360}.}
    \label{fig:ra_dec_scans}
\end{figure}

Figure \ref{fig:ra_dec_scans} shows the local probability around the location of the two most significant sources obtained from the point source search.

\section{Conclusion}\label{sec:conclusion}
Using the  flux upper limit obtained from the stacked search, we evaluate the contribution of blazars toward the total diffuse flux of astrophysical neutrinos using muon tracks reported by IceCube and at 100 TeV, it is at a maximum 7$\%$. From the results of this work, we cannot comment on the contribution from the non-blazar AGN or all hard X-ray AGN in the catalog due to the flux upper limits being non-constraining of the total neutrino flux from muon neutrino tracks.

The sources found in the point source analysis with the highest significance are Seyfert galaxies that are highly obscured. NGC 1068 has been reported in a previous IceCube study and hence, not considered in the final estimation of the significance evaluation. NGC 4151 is observed at a significance level of 2.9 $\sigma$.

From the stacking analysis, we do not find any evidence of significant neutrino emission from any of the hard X-ray AGN classes tested but  the second analysis where we searched for emission from individual sources resulted in two sources showing potential neutrino emission. Assuming the observed excess from the two sources as a neutrino signal, we see that their spectra are softer with  a best fit  spectral index of $\sim 3$  compared to the diffuse fluxes observed by IceCube with a spectral index closer to 2.5. This shows that the source class responsible for the bulk of the diffuse neutrino flux should have a harder spectrum than the sources found in this analysis. 



\bibliographystyle{ICRC}
\bibliography{references}

%

\clearpage

\section*{Full Author List: IceCube Collaboration}

\scriptsize
\noindent
R. Abbasi$^{17}$,
M. Ackermann$^{63}$,
J. Adams$^{18}$,
S. K. Agarwalla$^{40,\: 64}$,
J. A. Aguilar$^{12}$,
M. Ahlers$^{22}$,
J.M. Alameddine$^{23}$,
N. M. Amin$^{44}$,
K. Andeen$^{42}$,
G. Anton$^{26}$
% ...
\\
\\
% ...
$^{12}$ Universit{\'e} Libre de Bruxelles, Science Faculty CP230, B-1050 Brussels, Belgium \\
% ...
$^{17}$ Department of Physics, Loyola University Chicago, Chicago, IL 60660, USA \\
$^{18}$ Dept. of Physics and Astronomy, University of Canterbury, Private Bag 4800, Christchurch, New Zealand \\
% ...
$^{22}$ Niels Bohr Institute, University of Copenhagen, DK-2100 Copenhagen, Denmark \\
$^{23}$ Dept. of Physics, TU Dortmund University, D-44221 Dortmund, Germany \\
% ...
$^{26}$ Erlangen Centre for Astroparticle Physics, Friedrich-Alexander-Universit{\"a}t Erlangen-N{\"u}rnberg, D-91058 Erlangen, Germany \\
% ...
$^{40}$ Dept. of Physics and Wisconsin IceCube Particle Astrophysics Center, University of Wisconsin{\textendash}Madison, Madison, WI 53706, USA \\
% ...
$^{42}$ Department of Physics, Marquette University, Milwaukee, WI, 53201, USA \\
% ...
$^{44}$ Bartol Research Institute and Dept. of Physics and Astronomy, University of Delaware, Newark, DE 19716, USA \\
% ...
$^{63}$ Deutsches Elektronen-Synchrotron DESY, Platanenallee 6, 15738 Zeuthen, Germany  \\
$^{64}$ Institute of Physics, Sachivalaya Marg, Sainik School Post, Bhubaneswar 751005, India \\
% ...

\subsection*{Acknowledgements}

\noindent
The authors gratefully acknowledge the support from the following agencies and institutions:
USA {\textendash} U.S. National Science Foundation-Office of Polar Programs,
U.S. National Science Foundation-Physics Division,
U.S. National Science Foundation-EPSCoR,
Wisconsin Alumni Research Foundation,
Center for High Throughput Computing (CHTC) at the University of Wisconsin{\textendash}Madison,
Open Science Grid (OSG),
Advanced Cyberinfrastructure Coordination Ecosystem: Services {\&} Support (ACCESS),
Frontera computing project at the Texas Advanced Computing Center,
U.S. Department of Energy-National Energy Research Scientific Computing Center,
Particle astrophysics research computing center at the University of Maryland,
Institute for Cyber-Enabled Research at Michigan State University,
and Astroparticle physics computational facility at Marquette University;
Belgium {\textendash} Funds for Scientific Research (FRS-FNRS and FWO),
FWO Odysseus and Big Science programmes,
and Belgian Federal Science Policy Office (Belspo);
Germany {\textendash} Bundesministerium f{\"u}r Bildung und Forschung (BMBF),
Deutsche Forschungsgemeinschaft (DFG),
Helmholtz Alliance for Astroparticle Physics (HAP),
Initiative and Networking Fund of the Helmholtz Association,
Deutsches Elektronen Synchrotron (DESY),
and High Performance Computing cluster of the RWTH Aachen;
Sweden {\textendash} Swedish Research Council,
Swedish Polar Research Secretariat,
Swedish National Infrastructure for Computing (SNIC),
and Knut and Alice Wallenberg Foundation;
European Union {\textendash} EGI Advanced Computing for research;
Australia {\textendash} Australian Research Council;
Canada {\textendash} Natural Sciences and Engineering Research Council of Canada,
Calcul Qu{\'e}bec, Compute Ontario, Canada Foundation for Innovation, WestGrid, and Compute Canada;
Denmark {\textendash} Villum Fonden, Carlsberg Foundation, and European Commission;
New Zealand {\textendash} Marsden Fund;
Japan {\textendash} Japan Society for Promotion of Science (JSPS)
and Institute for Global Prominent Research (IGPR) of Chiba University;
Korea {\textendash} National Research Foundation of Korea (NRF);
Switzerland {\textendash} Swiss National Science Foundation (SNSF);
United Kingdom {\textendash} Department of Physics, University of Oxford.

\end{document}